\documentclass[journal=jpcbfk, manuscript=article, layout=onecolumn, articletitle=false]{achemso}
\pdfoutput=1

\usepackage[dvipsnames]{xcolor}
\usepackage{geometry}
\usepackage{graphicx}
\usepackage{graphics}
\usepackage[alsoload=synchem]{siunitx}
\DeclareSIUnit{\cal}{cal}
\usepackage{amsmath,amssymb}

\usepackage[stretch=15,shrink=15,step=3]{microtype}

\usepackage{hyperref}
\hypersetup{colorlinks, citecolor={blue},linkcolor={red}, urlcolor={blue},
pdftitle={The role of stacking in the dynamics of DNA hairpin formation},
pdfauthor={Majid Mosayebi, Jonathan P. K. Doye}, pdfdisplaydoctitle}

\newcommand{\etal}{{\textit {et al.~}}}

\title{The role of loop stacking in the dynamics of DNA hairpin formation}
\author{Majid Mosayebi}
\affiliation{Physical and Theoretical Chemistry Laboratory, Department of Chemistry, University of Oxford, South Parks Road, Oxford OX1 3QZ, United Kingdom}
\email{majid.mosayebi@chem.ox.ac.uk}
\author{Flavio Romano}
\affiliation{Physical and Theoretical Chemistry Laboratory, Department of Chemistry, University of Oxford, South Parks Road, Oxford OX1 3QZ, United Kingdom}
\author{Thomas E. Ouldridge}
\affiliation{Rudolf Peierls Centre for Theoretical Physics, 1 Keble Road, Oxford OX1 3NP, United Kingdom}
\author{Ard A. Louis}
\affiliation{Rudolf Peierls Centre for Theoretical Physics, 1 Keble Road, Oxford OX1 3NP, United Kingdom}
\author{Jonathan P.~K.~Doye}
\affiliation{Physical and Theoretical Chemistry Laboratory, Department of Chemistry, University of Oxford, South Parks Road, Oxford OX1 3QZ, United Kingdom}
\email{jonathan.doye@chem.ox.ac.uk}

\begin{document}


\begin{abstract}
We study the dynamics of DNA hairpin formation using oxDNA, a nucleotide-level coarse-grained model of DNA. In particular, we explore the effects of the  loop stacking interactions and non-native base pairing on the hairpin closing times. We find a non-monotonic variation of the hairpin closing time with temperature, in agreement with the experimental work of Wallace \etal [Proc. Nat. Acad. Sci. USA {\textbf{2001}}, {\textit {98}}, 5584-5589]. The hairpin closing process involves the formation of an initial nucleus of one or two bonds between the stems followed by a rapid zippering of the stem. At high temperatures, typically above the hairpin melting temperature, an effective negative activation enthalpy is observed because the nucleus has a lower enthalpy than the open state. By contrast, at low temperatures, the activation  enthalpy becomes positive mainly due to the increasing energetic cost of bending a loop that becomes increasingly highly stacked as the temperature decreases. We show that stacking must be very strong to induce this experimentally observed behavior, and that the existence of just a few weak stacking points along the loop can substantially suppress it. Non-native base pairs are observed to have only a small effect, slightly accelerating hairpin formation.
\end{abstract}

\section{Introduction}

Nucleic acid hairpins have diverse biological functions. For example, DNA hairpins play important roles in gene expression, DNA recombination and transposition\cite{DNA_bio4,DNA_bio1,DNA_bio2,DNA_bio3}. In RNA, hairpins are a common secondary structure motif and can serve as nucleating sites for higher order RNA structures\cite{RNA_hairpin_review}. In  addition, hairpins are commonly used in DNA-based nanotechnology, for instance as fuels in motors\cite{Green_hairpin, Eyal_motor}, as bio-sensors\cite{Tyagi_1996}, or for controlling self-assembly pathways\cite{assembly_Yin}. Together with duplex hybridization, hairpin formation is one of the most basic dynamical process involving nucleic acids and therefore a fundamental understanding of this process is of wide-ranging importance.

There have been many experimental and theoretical studies on the effects of different parameters such as the loop length, the sequence and the ionic strength of the solvent on the dynamics and thermodynamics of nucleic acid hairpins\cite{ansari_jpcb, libchaber_prl2000, Block_science,  eyal_hairpin, dorfman_hairpin, pande_hairpin, SalesPardo_model, stem_zipping,john_hairpin}. DNA hairpin formation is considered to be predominantly a two-state process with a rate-limiting step which involves loop closure and a subsequent rapid zippering of the stem\cite{ansari_biophysicsJ}. This hypothesis has been challenged by several recent studies that suggest a more complex pathway, involving intermediate states containing misfolded or partially folded configurations\cite{Orden_review, Orden_jacs, Ma_PEL}. Additionally, there have been diverse, and sometimes contradictory, experimental observations and interpretations of hairpin formation kinetics. Fluorescence energy transfer and fluorescence  correlation spectroscopy measurements\cite{libchaber_pnas} have reported an Arrhenius temperature dependency with a positive activation enthalpy for the hairpin closing rates. Other studies\cite{Wallace_2001, wallace_jpcb, ansari_jpcb} have reported a non-monotonic variation of the closing time with temperature and an effective activation enthalpy that switches sign from positive to negative close to the hairpin melting temperature. The reason for the positive activation enthalpy at low tempera$\tau$ure has been posited to be the slow configurational diffusion of the single-stranded DNA (ssDNA) due to the intrachain interactions\cite{Wallace_2001, ansari_jpcb, ansari_biophysicsJ}. For example, in Ref.\citenum{ansari_biophysicsJ}, it was argued that the intrachain interactions, including non-native WC bonds and also misstacked bases in the loop, increase the ruggedness of the free-energy surface at the early stages of hairpin formation, which in turn decreases the configurational diffusion coefficient and makes it significantly temperature dependent. An alternative explanation is that the positive activation enthalpy is a result of an increase in the ssDNA stiffness due to increasingly strong stacking in the loop as the temperature is decreased\cite{libchaber_pnas}. Evidence for this position comes also from experiments on the possible effects of the loop sequence on hairpin thermodynamics and dynamics\cite{libchaber_prl2000}. These experiments found that the melting temperature of a hairpin with a more strongly stacking poly(dA) loop is lower than for the equivalent hairpin with a poly(dT) loop, and that the hairpin closing times are longer for the hairpin with the poly(dA) loop.

To investigate in detail the effects of base stacking and non-native Watson-Crick (WC) base pairing ({\it i.e.}\ misbonding) on the stability and dynamics of hairpins, we have performed  extensive simulations of the kinetics and thermodynamics of the hairpin studied by Wallace \etal\cite{Wallace_2001}, using oxDNA\cite{tom_thesis, tom_model_jcp, petr_seq_dep}, a coarse-grained model at the nucleotide level that has been shown to be capable of describing single- and double-stranded DNA\cite{oxDNA_review} and  basic processes such as hybridization\cite{tom_hybridization} and toe-hold mediated strand displacement\cite{tom_toehold}. Importantly, oxDNA incorporates single-stranded stacking interactions and allows for the formation of non-native base pairs (base pairs that are not intended to be present in the stem). We first investigate whether oxDNA reproduces the non-monotonic dependence of hairpin closing times on temperature\cite{Wallace_2001}, and explore the role of stacking and non-native base pairs in the observed behaviour. We then consider perturbed stacking strengths to establish the sensitivity of hairpin closing times to this physical parameter.

\section{Methods}

\subsection{Coarse-grained model}
\begin{figure}
 \includegraphics[width=0.4\textwidth]{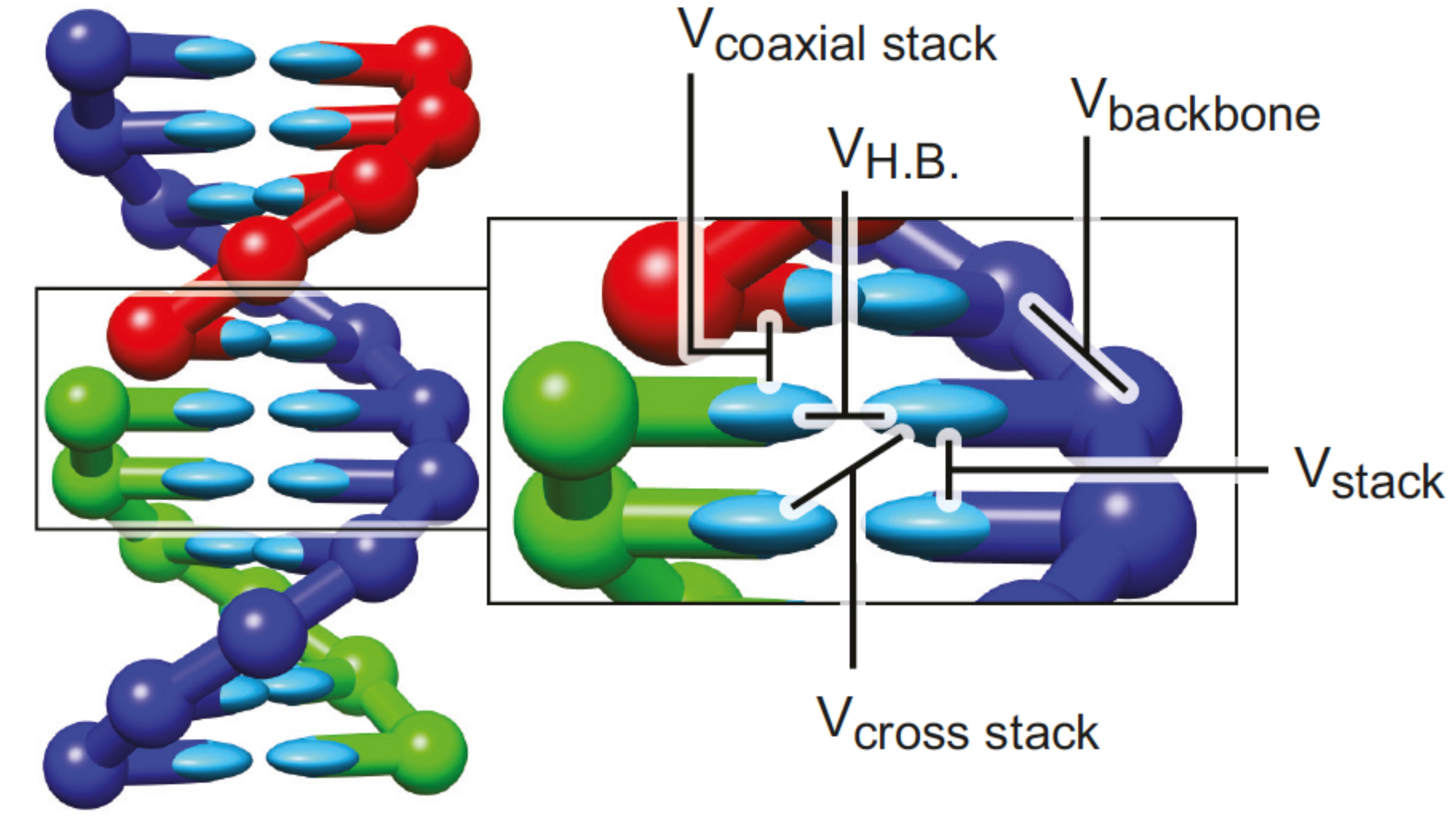}
 \caption{Illustration of a DNA double helix in the oxDNA model and the different interaction terms that stabilise the structure. The bases are represented by cyan ellipsoids and the backbone sites by spheres.}
 \label{fig:model}
\end{figure}

In oxDNA, each nucleotide is modelled as a rigid body with three interaction sites (see Figure \ref{fig:model}). The interaction potential consists of terms representing the backbone connectivity, excluded volume, hydrogen bonding between WC complementary base pairs, stacking between adjacent bases along the chain, coaxial stacking between non-adjacent bases, and cross stacking. Aside from backbone connectivity and excluded volume, all interactions are anisotropic, depending on the relative orientation of the nucleotides. The interactions are designed to favor the formation of a right-handed double helix at low temperature. The model has been previously described in detail\cite{tom_thesis, tom_model_jcp, petr_seq_dep} and is implemented in a simulation package which is available for download\cite{oxDNA}. The model has been parameterized for a salt concentration of \SI{0.5}{\Molar}, where the Debye screening length is short and it is reasonable to incorporate the electrostatic interactions in a soft excluded volume. This is the regime where most DNA nanotechnological experiments are carried out.  Here we use the  sequence-dependent parametrization of the model\cite{petr_seq_dep}, where the strengths of hydrogen bonding and stacking interactions depend on the identities of the interacting bases. 

OxDNA is particularly suited to the present study as it has been designed to reproduce the thermodynamic and mechanical properties of both single- and double-stranded DNA\cite{tom_model_jcp, petr_seq_dep}. As oxDNA incorporates single-stranded stacking interactions, and allows for the formation of non-native base pairs, simulating the model allows us to explore the consequences of these physical aspects of DNA for hairpin formation kinetics. In addition, oxDNA has been shown to accurately reproduce the most important aspects of  the kinetics of DNA duplex hybridization\cite{tom_hybridization} and toehold-mediated DNA strand displacement\cite{tom_toehold}. OxDNA has also been successfully applied to study a variety of DNA biophysical properties such as duplex overstretching\cite{flavio_overstretching} and plectoneme formation\cite{plectoneme}, as well as DNA nanotechnology systems, such as nano-tweezers\cite{tom_tweezers} and walkers\cite{tom_walker}.

\subsection{Simulation methods}

We perform dynamics simulations, where the mass, energy and length units are chosen to be $m_0=315.75$~\SI{}{\dalton}, $l_0= 8.518$~\SI{}{\angstrom}~and $\epsilon_0 = 4.142 \times 10^{-20}$~\SI{}{\joule},   respectively, implying a time unit $\tau_0=(m_0 l_0^2 / \epsilon_0) ^ {-1/2} = 3.03 \times 10^{-12}$~\SI{}{\second}. Each nucleotide has a mass of $m_0$ and a moment of inertia $I=0.138 ~ m_0l_0^2$. We use a weakly-coupled Andersen-like thermostat as implemented by Russo \etal\cite{john_thermo}~: the system is evolved by solving Newton's equation of motion using a velocity Verlet algorithm for $\approx 100$ steps with an integration time step of $dt=0.005 \tau_0$, and then the velocities and angular velocities of each particle are updated  from a Maxwell-Boltzmann distribution at the given temperature $T$ with probabilities $p_v=0.02$ and $p_\omega=0.0067$, respectively. On time scales much longer than $5000$ steps, the thermostat produces diffusive motion.

We study the kinetics of hairpin formation by performing brute-force closing simulations, as well as more complex simulations using the Forward Flux Sampling (FFS)\cite{FFS} method, which is a rare event method for accelerating kinetic measurements. It should be noted that, as is common for most coarse-grained models, it is not straightforward to map absolute time scales in our model onto experiment\cite{oxDNA_review}. We expect, however, that the relative time scales of similar processes in oxDNA to be directly comparable with experiments, as has been shown recently for duplex hybridization\cite{tom_hybridization,john_hairpin} and strand displacement\cite{tom_toehold}.

We obtain free-energy profiles using the Virtual Move Monte Carlo (VMMC) algorithm of Whitelam and Geissler\cite{Whitelam_VMMC}. We have found that VMMC greatly improves the equilibration speed  in oxDNA\cite{tom_thesis}. In addition, in order to efficiently sample the free-energy landscapes of these hairpin systems, we use the umbrella sampling technique\cite{us}. Umbrella sampling allows the system to overcome free-energy barriers by artificially biasing the system to sample states with higher free energy more frequently. 

Base stacking and hydrogen bonding between complementary base pairs are the two crucial interactions that cause DNA to behave differently from a normal polymer chain. Here, to investigate the role of stacking interactions on the dynamics of hairpin formation, we perform additional simulations in which all stacking interactions in the loop are scaled by a factor $\lambda_{\rm st}$ relative to their (sequence-dependent) values from Ref. \citenum{petr_seq_dep}; $\lambda_{\rm st}=1$ therefore corresponds to the canonical values quoted in Ref. \citenum{petr_seq_dep}.  To clarify the role of misbonding, we also simulate hairpins in which the complementary hydrogen-bonding interactions are switched off either (i) completely for all non-native base pairs, or (ii) just for non-native base pairs where at least one base belongs to the loop. In the latter case, the hairpin can still form misbonded base pairs in the stem.

\section{Results}
We initially consider two hairpins, each with a 30-base loop and a 5 base-pair stem. The sequences that we study are:
\begin{itemize}
\footnotesize
\item 
$(S_1) ~~3^\prime-CCCAA~ (A)_{30 }~TTGGG-5^\prime$
\item 
$(S_2) ~~3^\prime-CGCTA ~(A)_{30} ~TAGCG-5^\prime,$
\end{itemize}
in which the sequence $S_1$ is chosen to be the same as the hairpin studied by Wallace \etal\cite{Wallace_2001}.  To study the effect of misbonding on the kinetics, we also consider the second hairpin $S_2$, that is  very similar to $S_1$ but with a slightly altered sequence in order to reduce the amount of misbonding between the two stems. While the hairpin $S_1$ is able to make $8$ different stem-stem misbonded base pairs, there are only two misbonding possibilities available to $S_2$.

\subsection{Hairpin closing time}\label{closing}

\begin{figure}
  \centering
  \includegraphics[width=0.6\textwidth]{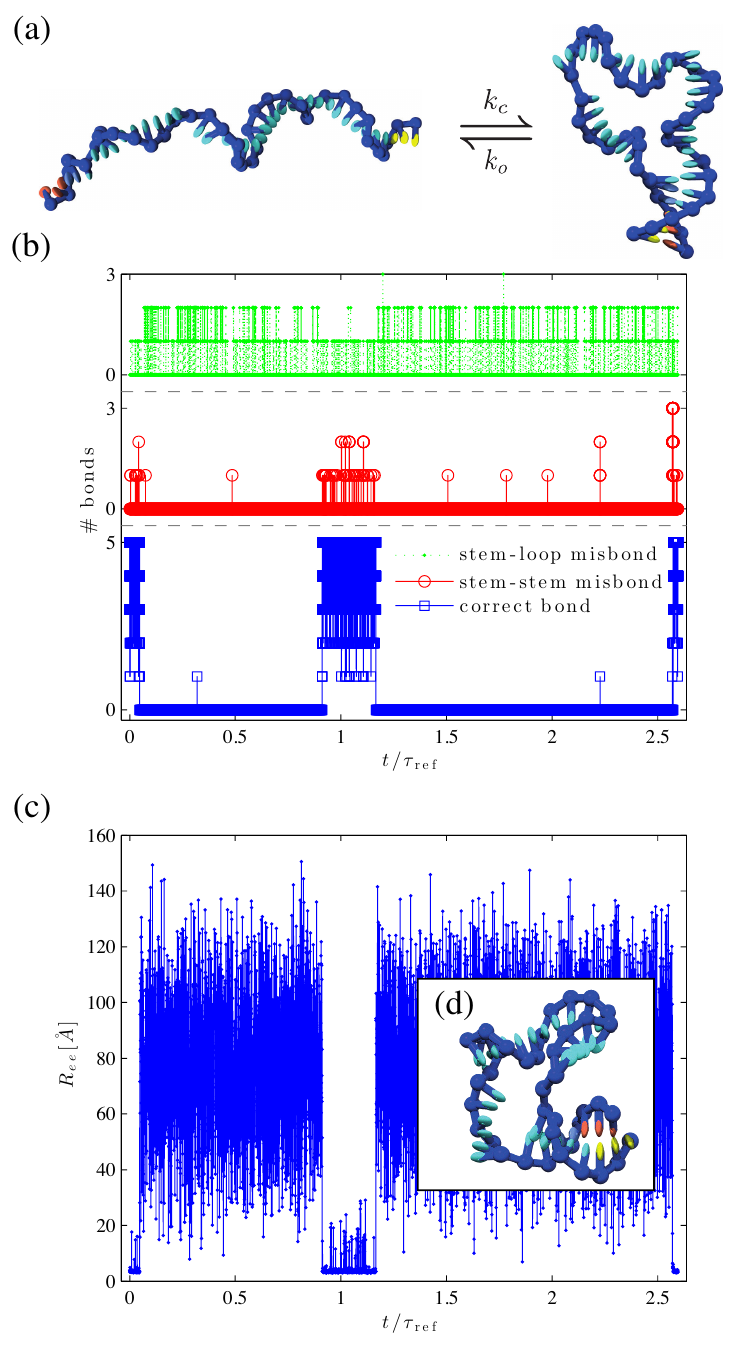}
 \caption{(a) Typical snapshots of the $S_1$ hairpin at $T=300$ \SI{}{\kelvin} in open and closed states. 
 (b) Number of base pairs and (c) the  end-to-end distance as a function of time at $T=310$ \SI{}{\kelvin}. Two different closing pathways occur. In the first closing event at $t/\tau_{\rm ref} \approx 0.91$, the hairpin folds by first forming correct base pairs in the stem, while in the second at $t/\tau_{\rm ref} \approx 2.57$ an initial stable closed loop is formed by misbonded base pairs which then rearrange to form a complete hairpin. 
 (d) $S_1$ hairpin with two stem-stem misbonds and two stem-loop misbonds. In (b) and (c) the time is normalized with respect to $\tau_{\rm ref}$, the average closing time of the $S_1$ hairpin at $T=300$ \SI{}{\kelvin}.
}
 \label{fig:closing_event}
\end{figure}

Figure \ref{fig:closing_event} illustrates typical configurations for the open, closed and misbonded states as well as the temporal evolution of the number of base pairs (a base pair being defined by a hydrogen-bonding energy of less than \SI{-0.596}{\kilo\cal\per\mol}) and the end-to-end distance $R_{ee}$ for the hairpin $S_1$. The open and closed states can be clearly distinguished. Furthermore, the transition from one state to the other is very fast compared to the time scale that the hairpin spends in each of the states. 

\begin{figure}
  \centering
  \includegraphics[width=\textwidth]{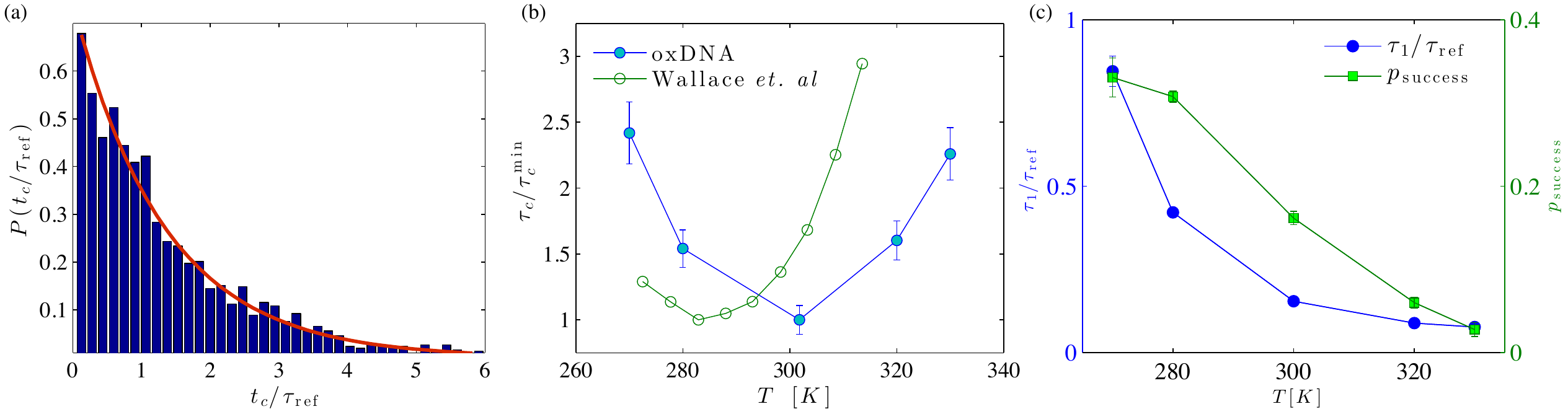}
 \caption{ 
 (a) Distribution of closing times $\tau_c/\tau_{\rm ref}$ for the $S_1$ hairpin at $T=280$~\SI{}{\kelvin}. Closing times are normalized with respect to $\tau_{\rm ref}$, the average closing time of the $S_1$ hairpin at $T=300$ \SI{}{\kelvin}.  The solid line is an exponential function $\tau_c^{-1} \exp (-t_c/\tau_c)$.
 (b) Normalized closing time $\tau_c$ as a function of temperature for the $S_1$ hairpin simulated with oxDNA at a salt concentration of \SI{0.5}{\Molar} (filled circles), and the experimental results of Ref. \citenum{Wallace_2001} for the same hairpin but at a salt concentration of \SI{0.1}{\Molar} (open circles). 
 (c) First contact formation time $\tau_1$, and success probability $P_{\rm success}$ for the $S_1$ hairpin, where $P_{\rm success}$ is defined as  the probability that a loop with one base pair leads to a fully formed hairpin before opening.
}
 \label{fig:ClosingTime}
\end{figure}

For hairpin formation, we define the average closing time $\tau_c$ as the average time it takes for an open hairpin to form all its native base pairs in the stem for the first time. The open hairpin belongs to an ensemble of equilibrated configurations with no base pairs. The distribution of closing times for the $S_1$ hairpin at $T=280$~\SI{}{\kelvin} is shown in Figure~\ref{fig:ClosingTime}(a). The distribution follows an exponential form, with a characteristic time scale that matches the average closing time $\tau_c$ of the hairpin at $T=280$~\SI{}{\kelvin}. The single exponential distribution that we observe for $\tau_c$ is typical of two-state reactions, where the dynamics are governed by transitions between two well-defined states ({\it{i.e.}} the open and closed states) and for which there exists a well-defined closing rate constant $k_c = 1 / \tau_c$.

The kinetics of simple two-state reactions are expected to follow the Arrhenius law, in which the reaction rate constant $k_c$ is exponentially related to a temperature-independent activation enthalpy $H_a$ through the relation $k_c \propto \exp (- H_a / k_BT),$ where $k_B$ is the Boltzmann constant. Figure~\ref{fig:ClosingTime}(b) shows the closing time for the $S_1$ hairpin as a function of temperature (filled circles). This hairpin forms most efficiently at $T\approx300$~\SI{}{\kelvin}. $\tau_c$ has a minimum at this temperature and increases on both raising or lowering the temperature. The existence of this minimum  clearly  shows that the formation of this hairpin is a non-Arrhenius process with an apparent activation enthalpy $H_a=\rm d  (\ln \tau_c) / \rm d (1/T)$ that changes sign at $T\approx300$~\SI{}{\kelvin} and becomes larger in magnitude as the temperature deviates more from $T\approx300$~\SI{}{\kelvin}. This observation is in qualitative agreement with the experimental measurements of Ref. \citenum{Wallace_2001} on the same hairpin (see Figure~\ref{fig:ClosingTime}(b)). Quantitative agreement with the experiment of Ref. \citenum{Wallace_2001} is difficult to achieve because oxDNA is parameterized at a higher salt concentration, and the rates are sensitive to this. Nevertheless, the changes we measure in our relative rates are of a similar order to those measured in the experiments \cite{Wallace_2001, wallace_jpcb, ansari_pnas, ansari_jpcb}. In addition, the overall shift of the oxDNA curve to higher temperatures compared to the experimental curve may in part reflect the further stabilization of the $S_1$ hairpin at higher salt concentrations. For example, the relative melting temperatures are \SI{305}{\kelvin} at \SI{0.5}{\Molar} in oxDNA and \SI{291.6}{\kelvin} at \SI{0.1}{\Molar} in the experiment. We note that duplex hybridization in the oxDNA model also exhibits non-Arrhenius behaviour, where the apparent activation enthalpy, although always negative, increases in magnitude with increasing temperature\cite{tom_hybridization}. 

To better understand the non-Arrhenius behaviour associated with the hairpin formation, we write the closing time as $\tau_c \approx \tau_1 /  P_{\rm success}$, where $\tau_1$ is the average time it takes to form the first stem-stem base pair with a hydrogen-bonding energy of less than \SI{-0.596}{\kilo\cal\per\mol}, and $P_{\rm success}$ is the probability of successful formation of the hairpin ( {\it{i.e.}}, with all native base pairs in the stem) starting from a state with one stem-stem base pair before returning back to the open state. The values of $\tau_1$ and $P_{\rm success}$ are obtained using the FFS technique\cite{FFS}. We note that the basic assumption in the application of FFS to measure rates of hairpin formation is that the zippering of the stem is much faster than the loop closure. This assumption is valid for the hairpin studied here, since $\tau_c$ computed with brute-force simulations that do include the zippering time agrees with the $\tau_c$ obtained with the FFS method. Figure~\ref{fig:ClosingTime}(c) shows that the success probability $P_{\rm success}$ is significantly reduced when the temperature is raised. At high temperatures, a single base pair  is not sufficiently stable to ensure formation of the complete hairpin and most of the times the loop opens up before zippering of the rest of the stem occurs. The rate limiting step for the formation of the hairpin at those temperatures therefore involves the search for a state which has on average more than one base pair. Although hairpin formation is predominantly a two state process, it is characterized by a complex set of transition states that have on average a larger number of base pairs at higher temperatures. As these transition states are enthalpically more stable (due to the base pairing) than the open state, a negative activation enthalpy is observed at high temperatures (where $\tau_1$ is relatively temperature independent) whose magnitude increases with increasing temperature. Negative activation enthalpies were previously observed for hybridization of oxDNA duplexes for similar reasons \cite{tom_hybridization}.

On lowering the temperature, $P_{\rm success}$ increases until it reaches a plateau at very low temperatures. In this regime the  temperature dependence is dominated by the change in the average time it takes to form the first stem-stem base pair ({\textit i.e.}\ $\tau_1$). The rapid increase of $\tau_1$ with decreasing temperature causes the closing time to pass through a minimum and then increase again. Thus the apparent activation enthalpy becomes positive at low temperatures. The positive activation enthalpy can be caused by variation in the enthalpy of the open state as well as the enthalpy of transition state ensemble. We will explore the microscopic origins of the low-temperature rise of $\tau_c$ later.

\subsection{Thermodynamics of hairpin formation}

\begin{figure}
  \centering
    \includegraphics[width=0.5\textwidth]{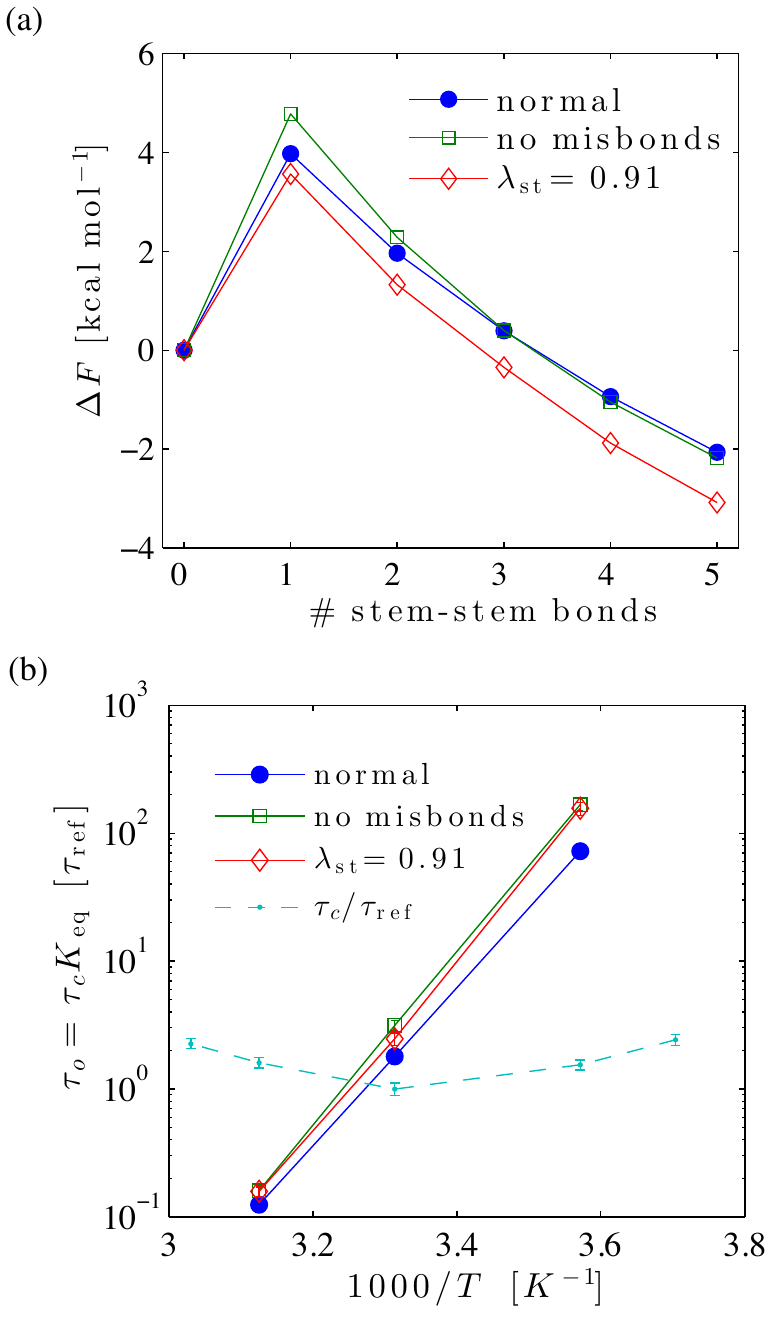}
  \caption{ (a) Free-energy profile as a function of the number stem-stem base pairs for the $S_1$ hairpin at $T=280$~\SI{}{\kelvin}. $\Delta F$ is measured relative to the open state with no stem-stem bonds. (b) Arrhenius plots of the opening time (solid lines), and closing time (dashed line) for the $S_1$ hairpin. $\tau_{\rm ref}$ is the average closing time of the $S_1$ hairpin at $T=300$ \SI{}{\kelvin}. In both (a) and (b) results are compared to when misbonding is switched off and when the stacking in the loop is reduced.}
 \label{fig:F}
\end{figure}

To gain further insight into the mechanism of hairpin formation we compute the free energies of the hairpin using the umbrella sampling technique.  In Figure \ref{fig:F}(a), we plot the free energy as a function of the number of stem-stem bonds at $T=280$ \SI{}{\kelvin}, showing that the $S_1$ hairpin is more stable than the open state by $\approx  2$ \SI{}{\kilo\cal\per\mole} at this temperature. 

For the hairpins that are studied here, the opening rates vary by several orders of magnitude within the temperature range of our study. At low temperatures, the hairpin opening process is slower than it could be conveniently measured with a brute-force method. Therefore, we calculate the opening time with help of the free-energy profiles that we obtain with umbrella sampling. In particular, we calculate the average opening time $\tau_o$ from the relation $K_{\rm eq} = \tau_o/ \tau_c$, where the equilibrium constant $K_{\rm eq}$ is the ratio of the partition functions of the closed (with at least one stem-stem base pair) and open states. The ratio can readily be obtained from free-energy profiles, such as those illustrated in Figure~\ref{fig:F}(a). The Arrhenius-plot in Figure~\ref{fig:F}(b) illustrates that, in contrast to hairpin formation, the opening of the $S_1$ hairpin exhibits Arrhenius behaviour in the temperature range of our study.  

\begin{figure}
  \centering
    \includegraphics[width=0.85\textwidth]{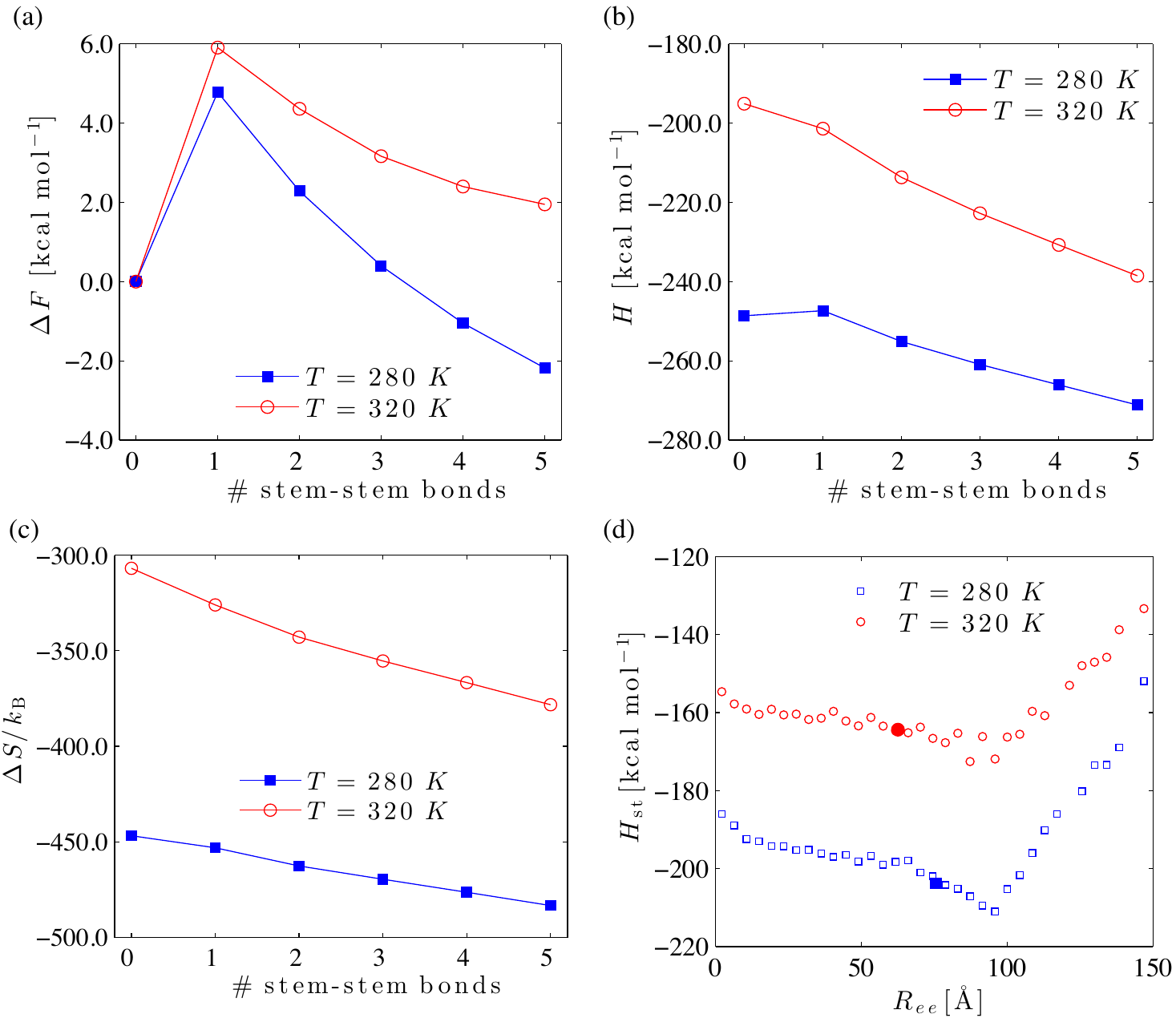}
  \caption{(a) Free energy, (b) enthalpy and (c) entropy of the $S_1$ hairpin as a function of the number of bonds in the stem at $T=280$ \SI{}{\kelvin} (solid square) and at $T=320$ \SI{}{\kelvin} (open circles) when misbonding is not permitted. (d) Stacking enthalpy $H_{\rm st}$ as a function of end-to-end distance $R_{\rm ee}$ for a 30-base poly(dA) ssDNA at two temperatures. Filled symbols mark the respective mean values $\langle H_{\rm st} \rangle$ and $\langle R_{\rm ee} \rangle$.}
 \label{fig:FHS}
\end{figure}

We showed that the apparent activation enthalpy of the formation of the $S_1$ hairpin inferred from the kinetic data is positive at low temperatures, implying that the enthalpy of the open state is lower than the enthalpy of the transition state. This is in contrast to the apparent negative activation enthalpy at high temperatures. In Figure \ref{fig:FHS}(b) we decompose the free-energy profiles at high and low temperatures into their enthalpic and entropic components when misbonding is not allowed. Misbonding is forbidden to simplify the analysis. We observe that the enthalpy difference between the state with one base pair and the open state is positive at $T=280$ \SI{}{\kelvin}, $ \Delta H = H(1)-H(0) \approx 1.28$ \SI{}{\kilo\cal\per\mole}. While the sign of $\Delta H$ is in agreement with our kinetic data, the apparent activation enthalpy $H_a=\frac{\rm d  \ln \tau_c }{\rm d (1/T)} \approx 3.33$ \SI{}{\kilo\cal\per\mole} is somewhat larger than $\Delta H$. This apparent discrepancy probably arises because the number of stem-stem base pairs is not a perfect reaction coordinate, {\it i.e.}, the transition state for hairpin formation at low temperatures does not coincide with an equilibrium population of states having one base pair by our energy criterion.

We can also see from Figure \ref{fig:FHS}(b) that at $T=320$ \SI{}{\kelvin}, the enthalpy difference $ \Delta H \approx - 6.28$ \SI{}{\kilo\cal\per\mole} is less negative than the apparent activation enthalpy $H_a\approx -7.19$ \SI{}{\kilo\cal\per\mole}, implying that the transition state has on average more than one base pair in this regime. Consistent with this argument, the success probability after forming one base pair is very small at $T=320$ \SI{}{\kelvin} (Figure \ref{fig:ClosingTime}(c)).

Hairpin loop closure of course involves a large loss of conformational entropy, but there is also an enthalpic cost to bringing the two ends of the chain together, because states with small $R_{\rm ee}$ are on average less stacked than those typical of the open state. From Figure \ref{fig:FHS}(d) we can see that this enthalpy cost is greater at low temperatures, when the bases are more strongly stacked and the loop is less flexible. Consistent with our results, a positive activation enthalpy, which increases with decreasing temperature,  has been reported experimentally for the end-to-end collisions of poly(dA) strands \cite{plaxco_ssDNA}. Breaking of stacking interactions to bring the two ends of the strand together is therefore the origin of the positive apparent activation enthalpy at low temperatures. Note that in this way the ssDNA in the loop region behaves differently from a freely-jointed chain, a commonly used model for ssDNA, for which the cost of bringing the chain ends together would be purely entropic.

\subsection{Role of misbonding} 
\label{misbonding}

Non-native base pairs have the potential both to enhance and to hinder the formation of a target hairpin. They can provide alternative pathways where misbonded base pairs form first, followed by an internal rearrangement to form the correct structure. Whether such pathways enhance the rate of hairpin formation depends on the ease with which misbonded structures can be resolved. For short duplexes, it has been previously shown that misbonding enhances the rate of hybridization because the time scale for internal displacement reactions that allow the correct duplex to be achieved after misbonding are much faster than the diffusion time associated with  base-pair forming encounters at typical (low) strand concentrations\cite{tom_hybridization}. However, whether a similar time scale separation holds for hairpin formation is less obvious because of the unimolecular nature of the process, and is likely to depend on the structure of the the hairpin ({\it e.g.} loop length), sequence, and parameters such as temperature that affect the ease with which free-energy barriers can be overcome. 

\begin{figure}
  \centering
    \includegraphics[width=\textwidth]{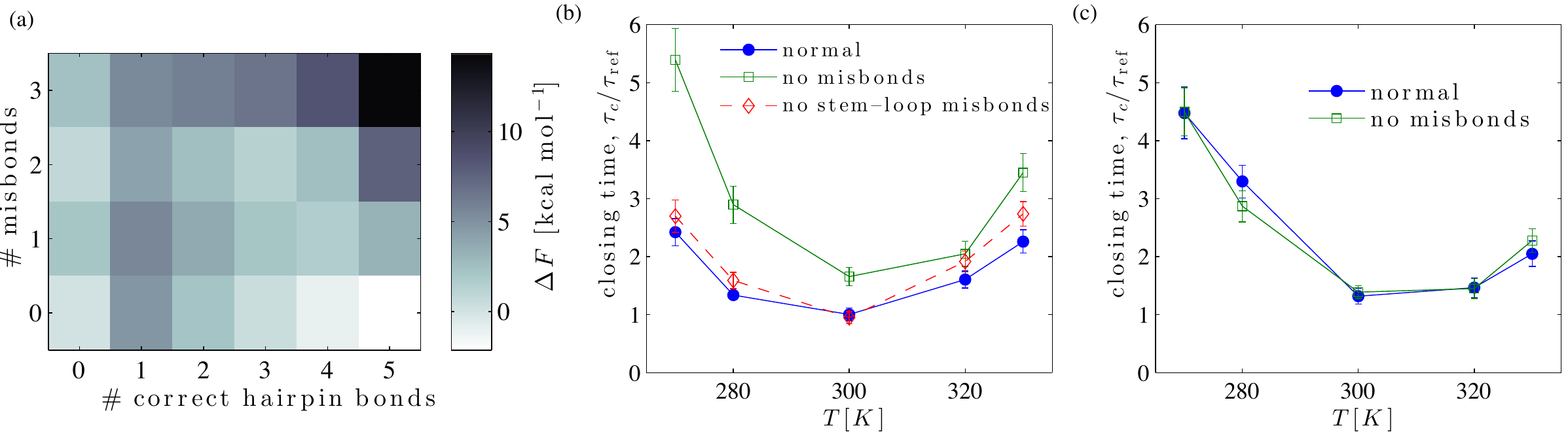}
  \caption{(a) 2-dimensional free-energy landscape showing the free energy of the $S_1$ hairpin at $T=280$~\SI{}{\kelvin} as a function of the number of misbonds and the number of correct hairpin bonds. $\Delta F$ is measured relative to the open state with no base pairs. The average closing time of (b) the $S_1$ hairpin and (c) the $S_2$ hairpin as a function of the temperature. In (b) and (c), the results are compared to those when misbonding is not permitted. All times are normalized with respect to $\tau_{\rm ref}$ the average closing time of the $S_1$ hairpin at $T=300$ \SI{}{\kelvin}.
 }
 \label{fig:misbond}
\end{figure}

A two-dimensional free-energy landscape as a function of the number of correct bonds in the stem and the total number of misbonds is plotted for the $S_1$ hairpin at $T=280$ \SI{}{\kelvin} in Figure \ref{fig:misbond}(a). The profile shows two minima corresponding to the fully open and closed states. In addition, we observe that various misbonded configurations  are possible. In particular, there is a local minimum involving two non-native WC base pairs (and no native base pairs). Moreover, the apparent free-energy barrier separating such a configuration from the fully-formed hairpin is actually smaller than the barrier of the direct path from the fully open state to the fully-formed hairpin. It is thus plausible that folding of the hairpin can occur through an alternative pathway initially involving non-native base pairs.

In kinetic measurements of the hairpin $S_1$, which has a short stem and a relatively long loop, we have considered the effects of completely switching off WC misbonding and only switching off WC misbonding between bases in the stem and in the loop. Figure~\ref{fig:misbond}(b)  shows that the closing time becomes longer, particularly at low temperatures, when misbonding is not allowed for the $S_1$ hairpin. However, stem-loop misbonds turn out to be less relevant in determining the closing time for this hairpin. It is interesting to note that for the hairpin $S_1$, misbonding in the stem significantly reduces $\tau_1$ (from $\approx1.25 \, \tau_{\rm ref}$ to $\approx0.42 \, \tau_{\rm ref}$ at $T=280$ \SI{}{\kelvin}), but it only reduces $P_{\rm success}$ slightly (from $\approx0.37$ to $\approx  0.31$ at $T=280$ \SI{}{\kelvin}) as the transient entrapment of the hairpin in misbonded states only makes the zippering stage of the hairpin formation slightly less likely. Note that the effect of misbonding that we observe here works against the the non-monotonic behaviour of $\tau_c$, rather than causing it. For hairpin $S_2$, on the other hand, there are fewer misbonding possibilities and therefore misbonding does not play an important role in the hairpin closing process, as is shown in Figure~\ref{fig:misbond}(c).

Figure~\ref{fig:F}(a) also compares the free-energy profiles for hairpin formation to the case when misbonding is not allowed. We observe that misbonding stabilizes the states with high free energy and therefore reduces the effective free-energy barriers of both closing and opening reactions. 
This explains, once more, why misbonding is assisting the hairpin formation/opening for the sequence $S_1$. 
Note that misbonds can only slow down hairpin formation at low temperature if the following conditions hold: (a) the misbonded configurations must be stable relative to the open state and (b) the misbonded configuration should be unable to easily transition to the folded state (eg. via internal displacement\cite{tom_hybridization}). In oxDNA, for the hairpins studied here, these conditions do not hold. It is also difficult to see how WC interactions could cause such behaviour for physical DNA.

\subsection{Role of stacking}

\begin{figure}
  \centering
    \includegraphics[width=0.5\textwidth]{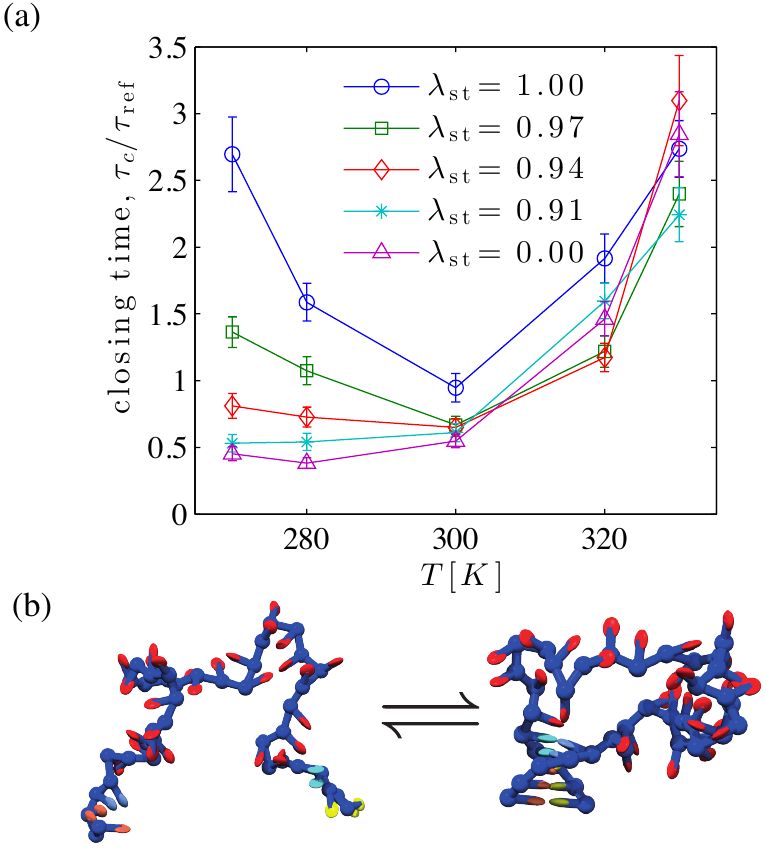}
 \caption{(a) Dependence of the closing time on $\lambda_{\rm st}$, the strength of base stacking in the loop. (b) Typical snapshots of the $S_1$ hairpin at $T=300$~\SI{}{\kelvin} in closed and open states when there is no base stacking in the loop, $\lambda_{\rm st}=0$.  }
  \label{fig:ClosingTime_vs_T_stacking}
\end{figure}

To further investigate the nature of the growth of $\tau_c$ at low temperatures, we systematically vary the stacking strength between the bases in the loop from no stacking $(\lambda_{\rm st} =0)$ to normal stacking $(\lambda_{\rm st} =1)$ and measure the closing time. The results are shown in Figure~\ref{fig:ClosingTime_vs_T_stacking}(a). Note that stem-loop misbonding is also forbidden, but this imposes only a negligible perturbation to the dynamics of the system, as was shown earlier. We observe that for large $\lambda_{\rm st}$ and at low $T$, where the bases in the loop stack strongly, the closing time grows. However, for $\lambda_{\rm st} \approx 0$ no such rise is observed.

The free-energy profiles in Figure~\ref{fig:F}(a) also indicate that  decreasing the stacking strength in the loop  stabilizes the closed states and thus shifts the melting point to a slightly higher temperature; such a shift was also observed in experiment when a loop sequence was replaced by one believed to stack less strongly\cite{libchaber_prl2000} and also studied previously for oxDNA\cite{petr_seq_dep}. The opening of the more stable hairpin is slower compared to the normal hairpin (see Figure \ref{fig:F}(b)). In addition, the reduction of the effective height of the free-energy barriers to hairpin formation that we see in Figure~\ref{fig:F}(b) further suggests that the rapid growth of the closing time observed at low temperatures is actually due to the strong stacking in the loop.

\begin{figure}
    \includegraphics[width=0.75\textwidth]{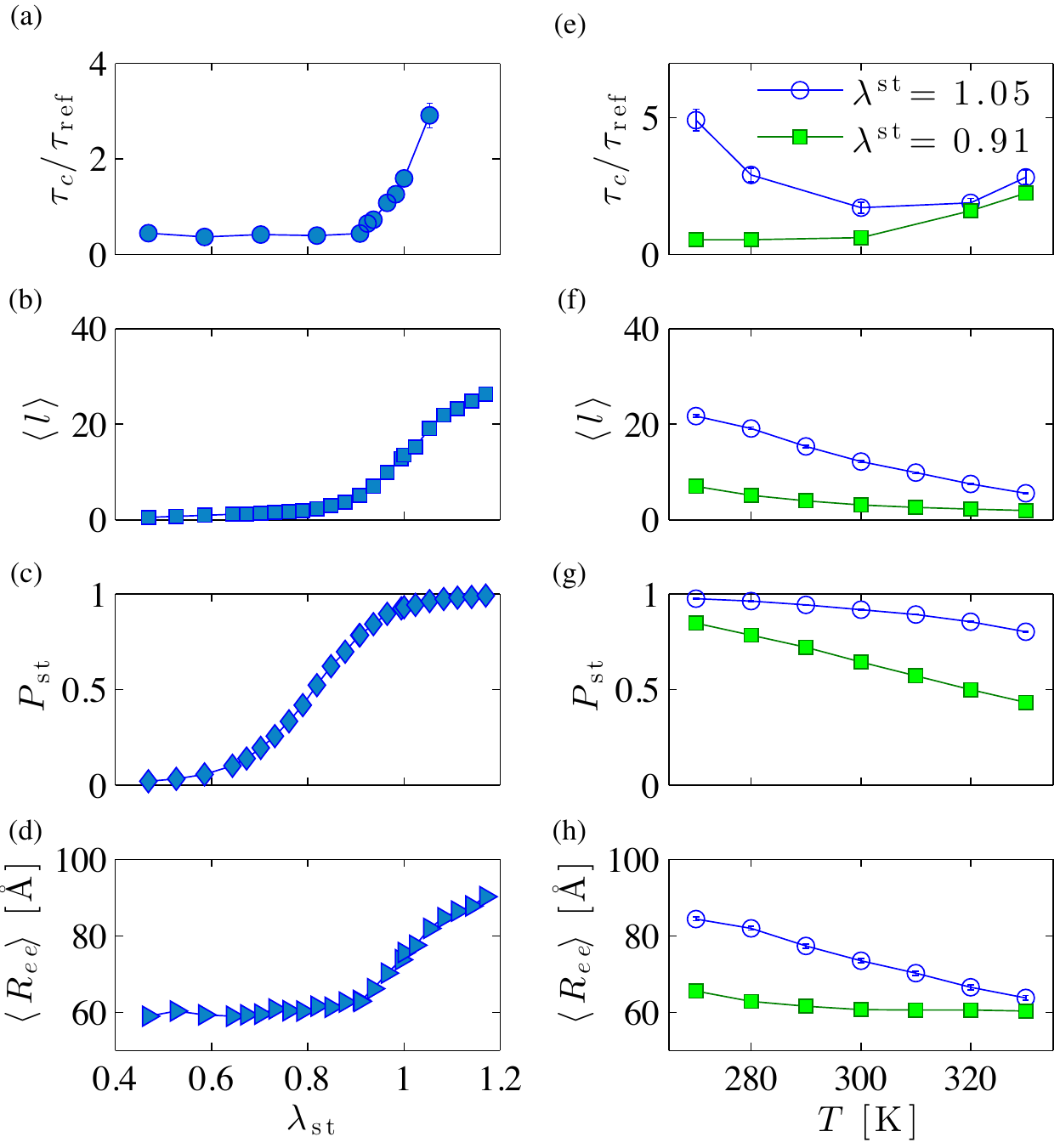}
 \caption{ Closing time $\tau_c$, average stacked length $\langle l \rangle$, probability of stacking $P_{\rm st}$, and end-to-end distance $R_{\rm ee}$ as a function of (a)-(d) the strength of stacking interaction when $T=280$\SI{}{\kelvin} and as a function of (e)-(h) the temperature when $\lambda_{\rm st} = 1.05$ (open circles) and when $\lambda_{\rm st} = 0.91$ (filled squares).	
 }
 \label{fig:stacking_length}
\end{figure}

To further quantify the effect of the stacking interactions, we measure the average stack length $\langle l \rangle$, the probability of stacking $P_{\rm st}$, and the average end-to-end distance $\langle R_{\rm ee} \rangle$ for a $30$-nucleotide poly(dA) ssDNA as the strength of AA stacking is varied. The stack length $\langle l \rangle$ is defined as the average over stacks of contiguously stacked bases (a stack of length $l$ consists of $l+1$ bases).  Two neighbouring bases are considered to be stacked if the stacking interaction between them is less than or equal to $\epsilon_c = -(0.596 \lambda_{\rm st}) $~\SI{}{\kilo\cal\per\mol}. 

The results as a function of temperature and stacking strength are summarized in Figure~\ref{fig:stacking_length}. For small  $P_{\rm st}$, the average stack length is close to zero and since there is little base stacking the DNA chain is very flexible. All quantities that are plotted in Figure~\ref{fig:stacking_length} remain almost constant  when $P_{\rm st} < 0.7$ and above this threshold they start to increase significantly. The stacking length for an infinite chain has previously been shown to behave  as   $\langle l \rangle =  P_{\rm st} / ( 1 - P_{\rm st} )$ in an un-cooperative model\cite{petr_seq_dep}. By lowering the temperature and/or increasing the stacking strength, $P_{\rm st}$, and therefore $\langle l \rangle$, increase. Note that in the strong stacking regime, due to the finite length of our ssDNA chain, $\langle l \rangle$ deviates from the above formula, which predicts $\langle l \rangle$ to diverge as $P_{\rm st} \rightarrow 1$. In this regime, the strong stacking of the bases leads to a stiff ssDNA chain. Comparing the plots of $\tau_c$ and $\langle l \rangle$ (Figure \ref{fig:stacking_length} (a),(b),(e) and (f) ) shows that the increase of $\tau_c$  for the hairpin $S_1$ at low temperatures coincides with the stiffening of ssDNA, whereas when the stacking is weaker $\langle l \rangle$ is always much shorter than the loop length and $\tau_c$ does not rise at low temperatures. It is interesting to note that the AA-stacking  is the strongest stacking interaction within oxDNA, and the consequences of slightly reducing its strength is to suppress the low-temperature rise of the closing time. Therefore, if stacking is the root cause of the observed behaviour in experiments \cite{Wallace_2001}, it must be very strong (corresponding to high stacking probabilities $\sim P_{\rm st}>0.9$).

\subsection{Sequence heterogeneity}

All the hairpins that we have studied so far have a homogeneous loop with identical bases. While the stacking among different bases does break and form over time (temporal heterogeneity) the average local stiffness of the ssDNA chain is homogeneous along the loop. Intuitively, one can imagine that introducing a few weak stacking points along the loop should effectively reduce the loop stiffness and therefore facilitate hairpin formation at low temperatures. To test this hypothesis, we replaced three of the adenine bases in the loop of the $S_1$ hairpin with ``dummy'' bases (D) for which we have chosen the strength of the stacking interactions to be $\lambda_{\rm st} ^{DA}=\lambda_{\rm st} ^{AD} = \lambda_{\rm st} ^{DD} = 0.91$ which would be similar in strength to our parameterisation of AT stacking in Ref. \citenum{petr_seq_dep}. Moreover, we vary the positions of the D bases in the loop to highlight any position-specific effects. The sequences that we consider are,
\begin{itemize}
\footnotesize
\item 
(I) ~~$3^\prime-CCCAA~ {\color {gray} D} AAAAAAAAAAAAA {\color {gray} D} AAAAAAAAAAAAA {\color {gray} D} A ~TTGGG-5^\prime$
\item 
(II) ~~$3^\prime-CCCAA ~AAAAAA {\color {gray} D}AAAAAAA {\color {gray} D} AAAAAAA {\color {gray} D} AAAAAAA ~TTGGG-5^\prime$
\item
(III) ~~$3^\prime-CCCAA ~AAAAAAAAAA {\color {gray} D} AAA {\color {gray} D} AAA {\color {gray} D} AAAAAAAAAAA ~TTGGG-5^\prime$.
\end{itemize}

\begin{figure}
\centering
  \includegraphics[width=0.5\textwidth]{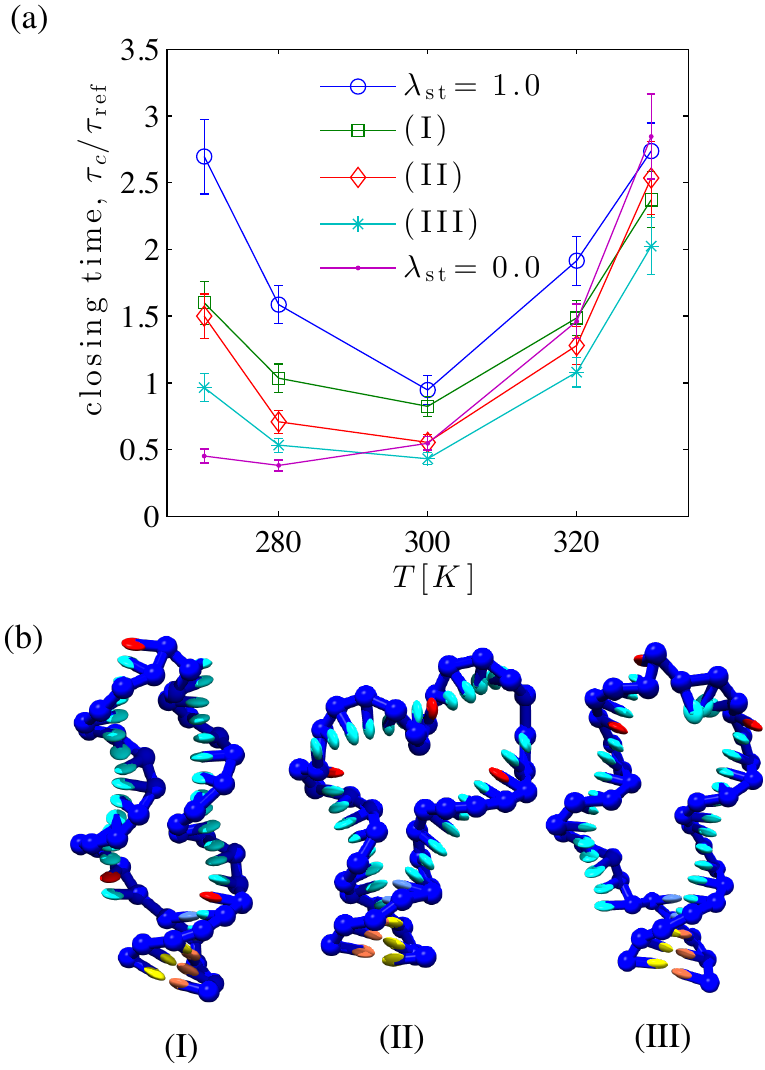}
  \caption{(a) Introduction of three weak stacking points along the loop significantly reduces $\tau_c$ at low temperatures. (b) Snapshots of the three types of the modified loop hairpins. Red colored bases mark the positions of the weak stacking points.}
 \label{fig:ClosingTime_vs_T_defect}
\end{figure}

Figure  \ref{fig:ClosingTime_vs_T_defect}(a) displays the closing time for the three modified hairpins compared to the normal $S_1$ hairpin and the hairpin with $\lambda_{\rm st}=0$. Our results show that the presence of just a few weak stacking points along the loop significantly affects the hairpin closing dynamics, making $\tau_c$ shorter. In addition, the positions of the weaker stacking points are important. We find that hairpin formation is easier when the weak stacking points are moved towards the middle of the loop, consistent with the results of Goddard \etal\cite{libchaber_prl2000} who found that replacement of an adenine base with a cytosine base in the middle of the loop of a similar hairpin was most effective at reducing the low-temperature closing time. For sequence III, where the weaker points are closer to the middle of the loop, the closing process is enhanced the most. For this hairpin, the three modified bases effectively reduce $\tau_c$ to a value comparable to the closing time of the hairpin with no base-stacking in the loop, except for the lowest temperature that we consider. These results are consistent with the intuitive expectation that the stem-forming ends of the strands are more likely to come close together if the strand can bend back on itself more easily at the middle of the hairpin loop.

\section{Discussion and conclusions} 

We have measured thermodynamic and kinetic properties of DNA hairpin formation with oxDNA, a coarse-grained model of DNA at the nucleotide level. The hairpins we consider, some of which have been studied experimentally, have a relatively long ($30$ bases) loop and a rather short ($5$ bases) stem. Our results indicate that the formation of such hairpins at high salt concentrations is well-approximated by a two-state reaction with a rate-determining step that involves the formation of a loop with a nucleus of one or two stem-stem base pairs. The formation of the hairpin is then achieved by a rapid zippering of the stem. In addition, the closing time shows a non-Arrhenius temperature dependence with a minimum close to the melting temperature, consistent with some previous experiments\cite{Wallace_2001, wallace_jpcb, ansari_pnas, ansari_jpcb}. 

At high temperatures, the temperature dependence of the closing time shows a negative apparent activation enthalpy ({\it i.e.} increases with temperature),  implying that the set of transition states have on average a lower enthalpy (due to base pairing) with respect to the open state. In this regime, the free-energy barrier to hairpin formation is mainly due to a significant loss of conformational entropy associated with the formation of the initial stem-stem contact. Moreover, similar to what has been recently observed for duplex hybridization\cite{tom_hybridization} with oxDNA, we found that at high temperatures a loop that is stabilized by just one base pair is unlikely to succeed in forming a complete hairpin. Instead, a larger number of stem-stem base pairs is required to make zippering most likely at high temperatures. Hence, the apparent activation enthalpy for the hairpin formation becomes more negative as the temperature increases, in agreement with several experiments\cite{Wallace_2001, ansari_pnas, ansari_jpcb} but in contrast with Ref.~\citenum{libchaber_pnas}.

The behavior changes at temperatures below the melting point. At these temperatures the closing time shows a positive apparent activation enthalpy ({\it i.e.} increases with decreasing temperature), consistent with several studies\cite{Wallace_2001, ansari_pnas, ansari_jpcb}. As the temperature is decreased, the bases in the loop stack more strongly together. To form an initial stem-stem contact at low temperatures requires the bending rigidity of the ssDNA loop to be overcome and involves a loss of stacking interactions, thus increasing the enthalpy of the transition states. At some point the enthalpic cost of forming a loop becomes larger than the negative enthalpy gain due to base pairing at the transition state, and as a result the apparent activation enthalpy becomes positive. The positive activation enthalpy has previously been attributed to the roughness of the free-energy surface, due to the misbonded (WC and non-WC misbonds including misstacked bases) configurations in the early stages of the formation of the loop\cite{ansari_pnas, ansari_biophysicsJ}. For oxDNA at least, we have shown here that this phenomenon is mainly due to strong base stacking in the loop and that the WC misbonding when present acts to suppresses the non-monotonic behaviour observed for the hairpin closing time.

In addition, we have shown that by decreasing the stacking strength of the bases in the loop, or by introducing a few more weakly stacking bases in the loop, the hairpin closing time decreases significantly at low temperatures, in agreement with experimental observations\cite{libchaber_prl2000}. The significance of introducing sites of enhanced flexibility for the hairpin dynamics provides further evidence for the key role of loop rigidity in determining the hairpin closing time at low temperatures. For example, although such a hairpin loop is more flexible than our standard case, the overall roughness of the free-energy landscape is likely to be almost the same for the two hairpins.

It should be noted that in our study we used an implicit solvent model. Therefore, the temperature dependency of the rate constants due to the change of the solvent viscosity is not captured. As the viscosity of the solvent decreases with increasing temperature, its effect would be  to move the minimum in the closing time to slightly higher temperature. Furthermore, hydrodynamic effects are neglected in our simulations. However, as all the strands that are considered here are the same length (40-bases), they are likely to experience similar hydrodynamic effects. Such effects are likely to be more relevant when studying the kinetics of hairpins with different sizes. 

Overall, our results suggest that strong stacking within the loop of a hairpin can produce hairpin folding kinetics consistent with experimental studies that have reported non-monotonic dependencies on temperature. We note that if neighbouring bases stack in a largely uncooperative fashion, the overall stacking strength must be very large (corresponding to high stacking probabilities $\sim P_{\rm st}>0.9$) to produce a strong signal. Such strong stacking is found in oxDNA for polyA at low temperatures. Whilst we do not find similar effects arising from non-native WC base pairs (since those misbonded configurations are not more stable than the open state, and they can easily transition to the folded state), there could possibly be other contributions to the observed behaviour. In particular, there may be non-WC interactions that are neglected in oxDNA but are sufficiently stable at low temperature to inhibit folding. Alternatively, single-stranded stem sections might form structures at low temperatures  that are incompatible with nucleating the correct stem and that must be disrupted prior to hairpin formation \cite{tom_hybridization}. Finally, changing interactions with the solvent or increased viscosity of water at lower temperatures may play a role. We note, however, that these alternative explanations are unlikely to be strongly sensitive to substitutions of a few bases within the loop, as is observed in oxDNA when strong stacking is responsible. The fact that such changes are known to influence folding dynamics at low temperature, therefore supports the stacking-based explanation rather than the suggestion of Ansari and coworkers \cite{ansari_pnas, ansari_biophysicsJ} that an increased roughness of the free-energy surface arising from interachain misbonding controls the hairpin formation dynamics.

\begin{acknowledgement}
The computational resources of the PolyHub virtual organization and Oxford's Advanced Research Computing are gratefully acknowledged. MM was supported by the the Swiss National Science Foundation (Grant No. PBEZP2-145981). The authors are also grateful to the EPSRC for financial support. 
\end{acknowledgement}

\providecommand{\latin}[1]{#1}
\providecommand*\mcitethebibliography{\thebibliography}
\csname @ifundefined\endcsname{endmcitethebibliography}
  {\let\endmcitethebibliography\endthebibliography}{}

\end{document}